\documentstyle[preprint,prl,aps]{revtex}

\begin{document}
\draft
\preprint{UCI-CMTHE-94-05}
\title{
The Doped Two Chain Hubbard Model
}
\author{ R.M.\ Noack and S.R.\ White }
\address{
Department of Physics
University of California
Irvine, CA 92717
}
\author{ D.J.\ Scalapino }
\address{
Department of Physics
University of California
Santa Barbara, CA 93106
}
\date{September 14, 1994}
\maketitle
\begin{abstract}
The properties of the two--chain Hubbard Model doped away from
half--filling are investigated.
The charge gap is found to vanish, but a finite spin gap exists over a
range of interchain hopping strength $t_\perp$.
In this range, there are modified $d_{x^2-y^2}$--like pairing correlations
whose strength is correlated with the size of the spin gap.
It is found that the pair field correlations are enhanced by the
onsite Coulomb interaction $U$.
\end{abstract}


\narrowtext

Studies of two chain $t$--$J$ and Hubbard models provide insight into
the development of pairing correlations in
strongly coupled electron systems which can exhibit a spin--gap phase.
For example, Lanczos calculations for a $t$--$J$ model by Dagotto et
al. \cite{dagotto} suggested that antiferromagnetic $S=1/2$ coupled
chains such as $(\text{VO})_2 \text{P}_2 \text{O}_7$ \cite{johnston}
should have a spin gap and that if this system could be doped, it
would have enhanced superconducting or charge--density--wave
correlations in the ground state.  Rice and coworkers
\cite{rice1,tsunetsugu} discussed
$\text{Sr}_2 \text{Cu}_4 \text{O}_6$,
which consists of layers containing weakly coupled $\text{Cu}
\text{O}$ ladders \cite{hiroi}, and based upon their analysis of a
$t$--$J$ model suggested that if this system were lightly doped, the
spin gap phase would persist and a ground state with dominant
superconducting correlations could be realized.

Various approaches to the two--chain Hubbard model have also been
reported \cite{oldnoack,yamaji,schulz,finkel,fabrizio,khveshchenko}.
%
%
%
Renormalization group calculations
\cite{schulz,finkel,fabrizio,khveshchenko} find evidence for a variety
of phases.  However, a number of these phases correspond to
renormalization flows to strong coupling, making it important to have
the guidance provided by independent numerical calculations.  Previous
density matrix numerical renormalization group calculations
\cite{noack} suggested that the two chain Hubbard model has a spin gap
at half--filling, provided the onsite Coulomb interaction $U$ and the
interchain hopping $t_\perp$ are finite.  For $U=8t$ and $t_\perp=t$,
with $t$ the one electron hopping along a chain, a spin gap was also
found for band fillings of $\langle n \rangle = 0.875$ and  $\langle n
\rangle = 0.75$ electrons per site.  In addition, for the doped case,
the equal time pair field correlations were observed to decay as a
power law, but with an exponent similar to that of the noninteracting
($U=0$) system.  Here we extend these two--chain Hubbard calculations
to explore the dependence of the doped state on the strength of the
interchain hopping $t_\perp$.  We find that the charge gap vanishes in
the doped state, but that a spin gap exists over a range of $t_\perp$.
Furthermore, the decay of the pair field correlations depends on
$t_\perp$, with the strongest correlations appearing when the spin gap
is a maximum.

The two chain Hubbard model has a Hamiltonian
\widetext
\begin{equation}
H=
- t \sum_{ i, \lambda \sigma}
( c^{\dagger}_{i,\lambda \sigma}c_{i+1,\lambda \sigma} +
 c^{\dagger}_{i+1,\lambda\sigma}c_{i,\lambda \sigma}  )
- t_\perp \sum_{i, \sigma}
( c^{\dagger}_{i,1 \sigma}c_{i,2 \sigma } +
 c^{\dagger}_{i, 2\sigma}c_{i, 1 \sigma} )
+ U\sum_{ i, \lambda}
n_{i,\lambda \uparrow}n_{i,\lambda \downarrow }.
\end{equation}
\narrowtext
Here $c^\dagger_{i,\lambda \sigma}$ creates an electron of spin
$\sigma$ on rung $j$ and chain $\lambda$, with $\lambda=1$ or 2.
The one-electron hopping along a chain is $t$, the hopping between
chains is $t_\perp$, and the onsite Coulomb interaction is $U$.
We will measure energies in units of $t$.

The calculations were carried out using the density matrix
renormalization group (DMRG) \cite{white}.  This technique allows the
calculation of ground state energies and correlation functions in a
controlled way.  The details of the techniques used for two Hubbard
chains are described in Ref. \cite{georgia}.
We studied lattices with open boundary conditions of various sizes up to
$2 \times 40$ sites, keeping up to 500 states per block.
The discarded weight of the density matrix (the truncation error) ranged
from $3.7 \times 10^{-5}$ to less than $10^{-8}$.  We estimate that
the maximum errors on the quantities shown in this paper are at most a
few percent, and typically are of the order of the plotting symbol
size or less.

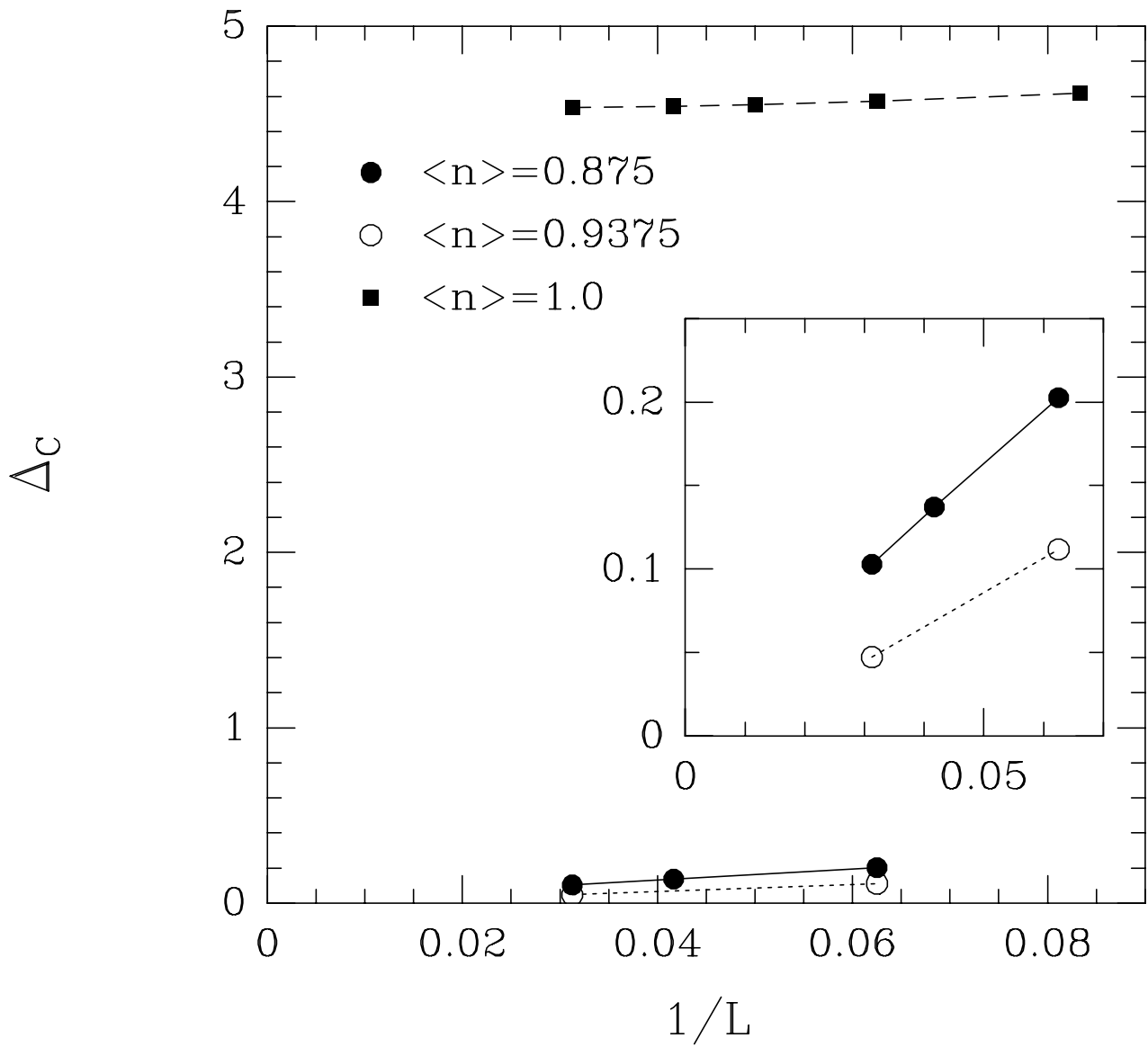
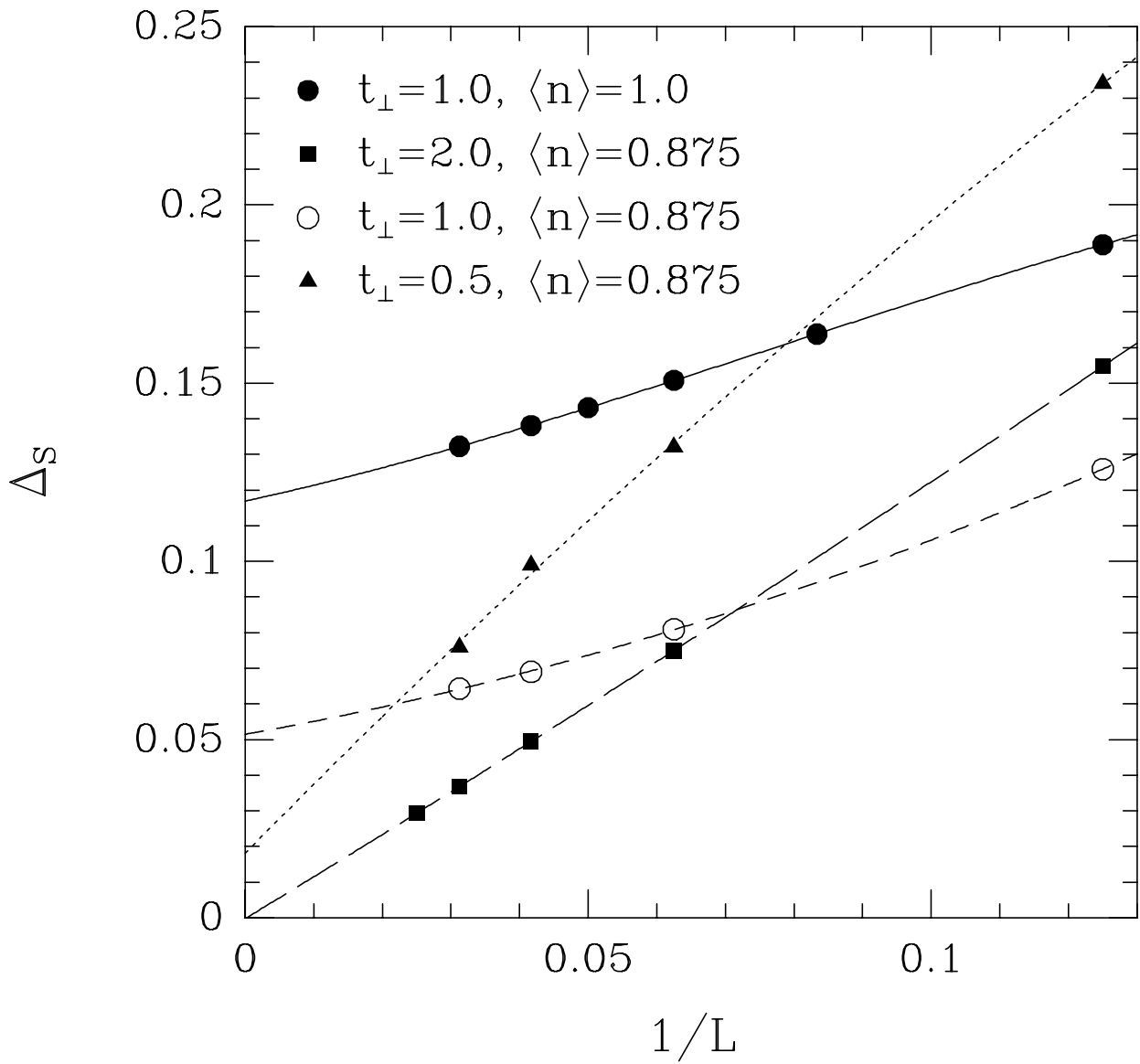
\begin{figure}
\caption{
(a) The charge gap $\Delta_C$ versus the inverse chain length $L^{-1}$
for $t_\perp=1$, $U=8$ and various fillings.
The inset shows an enlarged view for $\langle n \rangle = 0.875$.
(b) The spin gap $\Delta_S$ versus $L^{-1}$ for $t_\perp=1.0$ at
$\langle n \rangle =1.0$ and a number of $t_\perp$ values for $\langle
n \rangle=0.875$.
The lines are least squares fits to polynomials in $L^{-1}$.
}
\label{figgapl}
\end{figure}

The charge and spin gaps are defined in terms of the ground state
energy $E_0(N_\uparrow,N_\downarrow)$ for $N_\uparrow$ spin--up and
$N_\downarrow$ spin--down electrons.
The charge gap is given by
\begin{equation}
\Delta_C = [E_0(N-1,N-1) + E_0(N+1,N+1) - 2E_0(N,N)]/2,
\end{equation}
while the spin gap is
\begin{equation}
\Delta_S = E_0(N+1,N-1) - E_0(N,N).
\end{equation}
These quantities are shown in Fig. \ref{figgapl}, calculated for
Hubbard ladders of different lengths $L$.
In Fig. \ref{figgapl}(a) we show $\Delta_C$
for $U=8$, $t_\perp=1$, and several different band fillings.
At half--filling, $\Delta_C$ scales to a large finite value
$\Delta_C(L=\infty)\approx 4.5$, whereas
in the doped case, with $\langle n \rangle$ slightly
less than 1, we find that the charge gap vanishes in the thermodynamic
limit.
For the spin gap, shown Fig. \ref{figgapl}(b), the finite size scaling
must be done carefully in order to determine the behavior because
the finite size effects are relatively large and have a different
character in different regimes.
We fit the data using a least squares fit to polynomials in $L^{-1}$.
At half--filling, we use a third order polynomial to fit six system sizes.
At $\langle n \rangle = 0.875$, fewer sizes are available with the
correct filling, so we fit to a second order polynomial.
The polynomial fits are shown as the lines in \ref{figgapl}(b).
At half--filling, $\Delta_S(\infty)\approx 0.12$ and is clearly
nonvanishing.
The system is an insulator with short--range antiferromagnetic
correlations.
For the doped system, there are three distinct regimes for which we
show representative curves with different finite size behavior.
For small $t_\perp$, illustrated by the $t_\perp = 0.5$ curve, the finite
size effects are large with a large linear coefficient.
In the thermodynamic limit, the spin gap becomes small.
For a smaller $t_\perp$ of 0.3, we find
$\Delta_S(32)\approx 3 \times 10^{-3}$.
Therefore, we believe that $\Delta_S(\infty)$ vanishes for
$t_\perp < 0.5$.
For intermediate $t_\perp$, shown for $t_\perp =1.0$, the coefficient
of the quadratic term is large and positive and the spin gap scales to
a finite value.
For large $t_\perp$, shown for $t_\perp=2.0$, the finite size effects
are smaller, the scaling is linear in $1/L$ and $\Delta_S(\infty)$
vanishes.

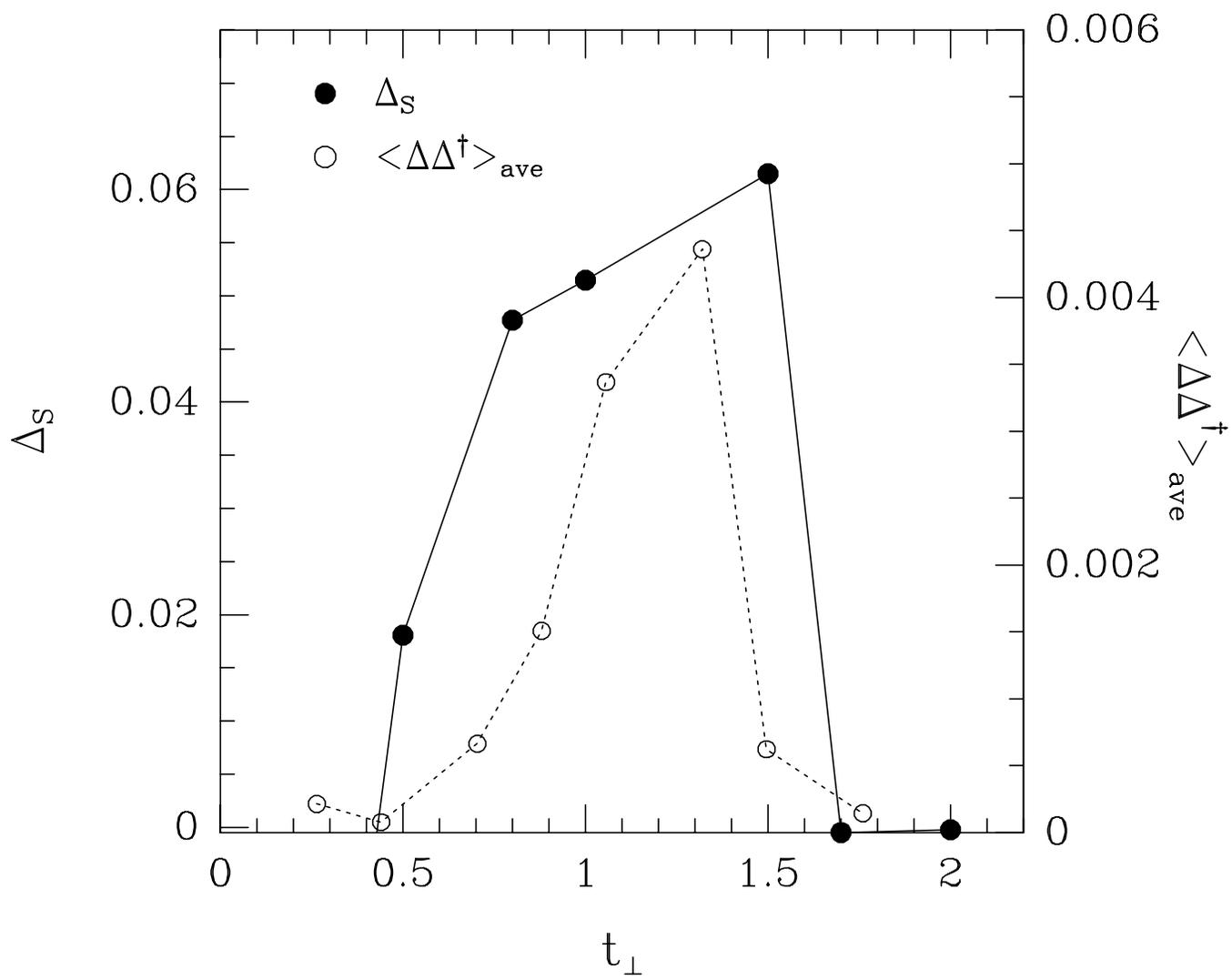
\begin{figure}
\caption{
The spin gap $\Delta_S$ versus $t_\perp$ for $U=8$ and
$\langle n \rangle = 0.875$ calculated from the $L\rightarrow \infty$
extrapolations of Fig. 1(b).
The open circles show the magnitude of the pair field correlation
function $\langle \Delta_i \Delta^\dagger_j \rangle$ averaged over separations
$8 \le |i-j| \le 12$ versus $t_\perp$ for the same parameters on a
$2 \times 32$ lattice.
}
\label{figgapt}
\end{figure}

We plot the spin gap extrapolated to the thermodynamic limit as a
function of $t_\perp$ in Fig. \ref{figgapt}.
It appears that in the doped system, a finite value of $t_\perp$ is
required to produce an interchain rung exchange $J_\perp \sim 4
t_\perp^2/U$ which is sufficient to create a spin gap.
This differs
from the half--filled case where our earlier work suggests that a spin
gap opens for all $t_\perp > 0$ \cite{noack,rvbprl}.
For $t_\perp > 1.7$,  as shown in Fig. \ref{figgapt}, the spin gap is
suppressed.
In a weak coupling picture, a two chain model leads to bonding and
antibonding bands split by the coupling $t_\perp$.
For $U=0$, the antibonding band is completely unoccupied at
$\langle n \rangle = 0.875$
when $t_\perp > 1.85$, suggesting that the
transition at $t_\perp \approx 1.7$ is related to the band splitting.
%
In the spin gap phase, the spin--spin correlations are short--range,
incommensurate, and antiferromagnetic in nature.  They give rise to a
Lorentzian peak in the magnetic structure factor \cite{noack}.

In order to further characterize the spin gap region, we have
calculated the equal--time rung--rung pair field correlation function
$\langle \Delta_i \Delta^\dagger_j \rangle$.
Here
\begin{equation}
\Delta^\dagger_j = ( c^\dagger_{j,1 \uparrow} c^\dagger_{ j,2 \downarrow}
- c^\dagger_{j,1 \downarrow} c^\dagger_{j,2 \uparrow})
\end{equation}
creates a spin singlet pair across the $j$th rung and $\Delta_i$
destroys a singlet pair on the $i$th rung.  For comparison with the
spin gap, Fig. \ref{figgapt} also shows the pair field correlation
function $\langle \Delta_i \Delta^\dagger_j \rangle$ averaged over
separations $8 \leq |i-j| \leq 12$.  It is clear that the size of the
spin gap and the pair field are correlated.

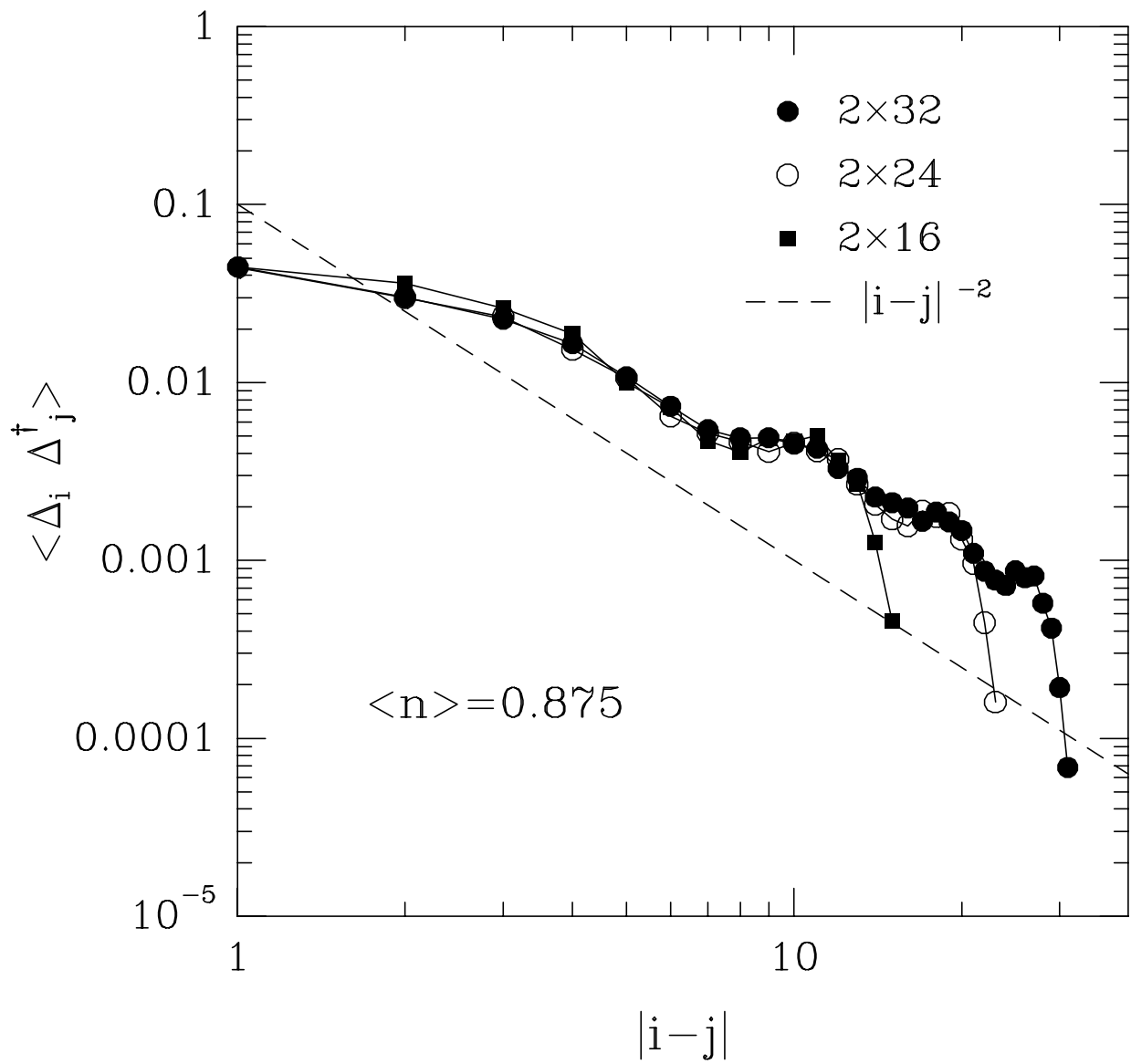
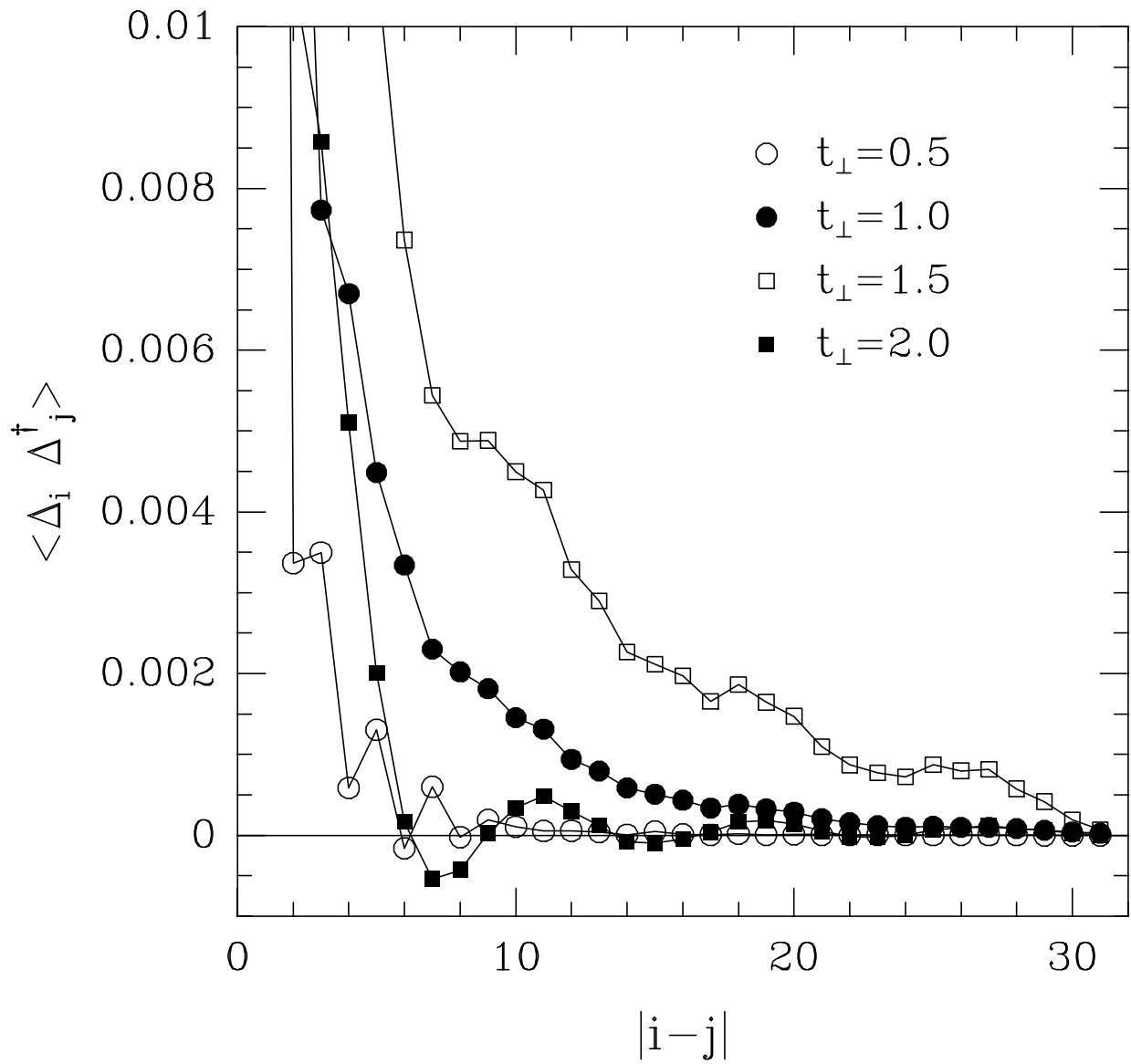
\begin{figure}
\caption{
(a) The equal time rung--rung pair field correlation function
$\langle \Delta_i \Delta^\dagger_j \rangle$ versus $|i-j|$ plotted
on a log-log scale for $t_\perp=1.5$ and for ladders of length $L=16$,
$L=24$, and $L=32$.
The dashed line corresponds to the $|i-j|^{-2}$ decay of the
non-interacting pair field correlations.
(b) The equal time pair field correlation function
$\langle \Delta_i \Delta^\dagger_j \rangle$ versus $|i-j|$ for
various values of $t_\perp$.
}
\label{figpair}
\end{figure}

Figure \ref{figpair}(a) shows a log--log plot of
$\langle \Delta_i \Delta^\dagger_j \rangle$ versus $|i-j|$ with
$\langle n \rangle = 0.875$, $U=8$ and $t_\perp =1.5$ for various
length chains.
As discussed earlier, the calculational procedure which we have
implemented has open end boundary conditions.
The correlation functions shown in Fig. \ref{figpair}(a) are
obtained by averaging over a number of $i$ and $j$ for a given
$|i-j|$.
We have found that this procedure removes most of the finite size
effects due to the open boundary conditions.
One can see that the boundary effects are small in
Fig. \ref{figpair}(a) by the overlap of the
numerical results for chains of different lengths.
Only when $|i-j|$ is large enough that both $i$ and $j$ are within a
few lattice spacings of the boundaries do significant differences
appear.
The dashed curve corresponds to the $|i-j|^{-2}$ decay of the
non--interacting pair field correlations
and one sees that for
$t_\perp=1.5$, the pair field correlations of the interacting system
are enhanced over the $U=0$ system.
In Fig. \ref{figpair}(b), we show $\langle \Delta_i \Delta^\dagger_j
\rangle$ versus $|i-j|$ calculated on a $2 \times 32$ system for
$\langle n \rangle = 0.875$, $U=8$, and various values of $t_\perp$.
For small values of $t_\perp$, the pair field correlations decay
rapidly.  As $t_\perp$ increases, the pair field correlations are
enhanced until $t_\perp \approx 1.7$, after which they are suppressed
by further increasing $t_\perp$ \cite{yamaji2}.

It is also possible to determine the internal structure of a pair.
We have calculated the pair wave function obtained from the
off--diagonal ground state expectation value
\begin{equation}
\Psi(i,j) = {}_{{}_{N-2}} \langle \psi_0 | (c_{j \uparrow} c_{i \downarrow}
- c_{j \downarrow} c_{i \uparrow}) | \psi_0 \rangle_{{}_{N}} .
\end{equation}
Here $(c_{j \uparrow} c_{i \downarrow} - c_{j \downarrow} c_{i \uparrow})$
removes a singlet pair with one electron on the $i$th
site and one on the $j$th site.
Near the center of the ladder, $\Psi(i+\hat{x},i) = 0.1$ and
$\Psi(i+\hat{y},i) = -0.08$ for $U=8$, $t_\perp=1.5$, and
$\langle n \rangle = 0.875$.
In the cuprate chain systems, $t_\perp$ and $t$ have the same sign so
that the relative negative sign of $\Psi(i+\hat{x},i)$ and
$\Psi(i+\hat{y},i)$ is physically relevant and corresponds to a modified
$d_{x^2-y^2}$--like state as previously discussed \cite{rice1,noack}.
The small amplitude for removing a pair from a rung, 0.1,
reflects the spatial extent of the relative, internal coordinate of
the pair wave function and the fact that the
$c^\dagger_{i,\lambda \sigma}$ operators are bare quasi--particle operators.
When one takes out a factor of $(0.1)^2$ associated with this overlap
of $\Delta_i$ and $\Delta^\dagger_j$ with the internal pair wave
function, it is clear that there are significant
pair-field--pair-field center of mass correlations present in the
doped system for $t_\perp=1.5$.

Recently, we described an intuitive picture for the spin--liquid state
of two coupled chains \cite{rvbprl} based on a resonating valence bond (RVB)
variational ansatz \cite{rvbpapers}.  This picture is also useful
for describing the apparent destruction of the spin--liquid state for
both small and large $t_\perp$. In the latter region, the bonding and
antibonding orbitals on a single rung separate to the extent that
a state with two holons on a rung becomes higher in energy than
two widely separated excitations each consisting of a hole in the bonding
orbital of a rung. These quasiparticle excitations are described in
RVB language as a bound holon-spinon pair, and were first observed by
Tsunetsugu, et. al. \cite{tsunetsugu}. The RVB picture suggests that
the spin--liquid state is not really destroyed in this regime, only
hidden by the spin degrees of freedom of the quasiparticles. This
picture is quite compatible with the weak--coupling, band--separation
picture mentioned above.

In contrast, a band picture is not useful for small $t_\perp$.  In
this regime, there is little difference in energy between the
``staggered'' and ``resonant'' valence bond
configurations\cite{rvbprl}, and the average separation between two
holons in a pair $\xi_h$ becomes substantial.  When $\xi_h$ becomes
comparable to $1/\delta$, where the doping $\delta=1-\langle n \rangle$,
the pairs overlap, and the holons become essentially free. Since they
are topological excitations residing on a single site, the
free holons destroy the confinement mechanism responsible for the
spin-liquid state. In this regime the system may resemble a doped
system with an odd number of chains, where confinement never takes
place.

It is interesting to compare our results with several renormalization
group calculations, which have also suggested the relevance of
$t_\perp$ and the possibility of interchain pairing
\cite{schulz,finkel,fabrizio,khveshchenko}.  Based upon a weak
coupling renormalization group analysis, Fabrizio et al.
\cite{fabrizio} have suggested a possible $U$--$t_\perp$ phase diagram
for the two chain Hubbard model.  In the small $t_\perp$ region,
they conjecture that there may be several phases in which pairing
correlations are dominant.  In one of these phases, called ``SC1'' in
Ref.  \cite{fabrizio}, these pairing correlations have the same
modified $d_{x^2-y^2}$ character we find.  However, contrary to our
results, they find a divergent spin--spin correlation function for
${\bf q} = (2 k_F, 0)$ rather than a spin gap state.  Furthermore,
they find this behavior for arbitrarily small values of $t_\perp$ and
for generic, noncommensurate fillings.  This differs from our findings
that the pairing is associated with a spin gap state and requires a
finite value of $t_\perp > 0.5$.  Furthermore, we find that a doping
of $\langle n \rangle$ near 1 is required for significant pairing.

Recently, Khveshchenko and Rice \cite{khveshchenko} have carried out a
renormalization group analysis near half--filling where umklapp
processes become relevant.   They argue that in the lightly doped
system, the presence of the interchain hopping $t_\perp$ is sufficient
to generate an antiferromagnetic exchange $J_\perp$ leading to a spin
gap, and that this state supports a modified $d_{x^2-y^2}$--like
pairing.  This scenario is certainly closer to what we find, except,
as shown in Fig. \ref{figgapt}, we find that a finite value of
$t_\perp$ is required to produce a spin gap.


The authors thank N. Bulut, M.P.A. Fisher, W. Hanke, M. Imada, T.M.
Rice, R.T. Scalettar, and H. Tsunetsugu for useful discussions.
R.M.N. and S.R.W. acknowledge support from the Office of Naval
Research under grant No. N00014-91-J-1143 and D.J.S. acknowledges support
from the Department of Energy under grant DE--FG03--85ER45197 and the
Program on Correlated Electrons at the Center for Materials Science at
Los Alamos National Laboratory.
The numerical calculations reported in this paper were performed at
the San Diego Supercomputer Center.

\end{document}